# A new concept for superior energy dissipation in hierarchical materials and structures

Giuseppe Puglisi[a], Nicola M. Pugno[b,c,∗]

[a] Dipartimento di Ingegneria Civile, Ambientale, del Territorio, Edile e di Chimica, Politecnico di Bari, Via Re David 200, Bari, Italy [b] Laboratory for Bioinspired, Bionic, Nano, Meta Materials & Mechanics, Department of Civil, Environmental and Mechanical Engineering, University of Trento, Italy
[c] School of Engineering and Materials Science, Queen Mary University of London, UK
∗ Corresponding author at: Laboratory for Bioinspired, Bionic, Nano, Meta Materials & Mechanics, Department of Civil, Environmental and Mechanical Engineering, University of Trento, Italy.
E-mail address:     nicola.pugno@unitn.it (N.M. Pugno).

ABSTRACT

We propose a new conceptual approach to reach unattained dissipative properties based on the friction of slender concentric sliding columns. We begin by searching for the optimal topology in the simplest telescopic system of two concentric columns. Interestingly, we obtain that the optimal shape parameters are material independent and scale invariant. Based on a multiscale self-similar reconstruction, we end-up with a theoretical optimal fractal limit system whose cross section resembles the classical Sierpiński triangle. Our optimal construction is finally completed by considering the possibility of a complete plane tessellation. The direct comparison of the dissipation per unit volume $\delta$ with the material dissipation up to the elastic limit $\delta el$ shows a great advantage: $\delta \sim 2000 \delta el$. Such result is already attained for a realistic case of three only scales of refinement leading almost (96%) the same dissipation of the fractal limit. We also show the possibility of easy recovering of the original configuration after dissipation and we believe that our schematic system can have interesting reliable applications in different technological fields.

Interestingly, our multiscale dissipative mechanism is reminiscent of similar strategies observed in nature as a result of bioadaptation such as in the archetypical cases of bone, nacre and spider silk. Even though other phenomena such as inelastic behavior and full tridimensional optimization are surely important in such biological systems, we believe that the suggested dissipation mechanism and scale invariance properties can give insight also in the hierarchical structures observed in important biological examples.

1. Introduction

Hierarchical structure design opened up the possibility of creating new devices with optimized mechanical properties (BührigPolaczek et al., 2016; Yong, Xiang, Jin, Chen, & Wang, 2018). Ultra-tough fibers have also been proposed in Pugno (2014) based on biomimetic design by introducing in the fiber a periodic sequence of sliders such as knots (see the extensions in Berardo, Pantano, and Pugno (2016), Bosia et al. (2016), Pantano, Berardo, and Pugno (2016)). Moreover, low scale tensegrity type systems, possibly multiscale, have been shown to deliver interesting possible applications for isolation based on their intrinsic strongly non linear geometrical properties (De Tommasi, Maddalena, Puglisi, & Trentadue, 2017; De Tommasi, Marano, Puglisi, & Trentadue, 2015; Trentadue, De Tommasi, & Marasciuolo, 2021). Also in nature, hierarchical structures have been recognized at the base of the capacity of dissipation and crush resistance of different biological systems such as bones (Currey, 2012), spider silks (Fazio, De Tommasi, Pugno, & Puglisi, 2022) and nacre (Oaki & Imai, 2005).

We propose a prototype of dissipative systems obtained by the composition of multiscale frictional devices arranged in a planar array, and characterized by a self-similarity of the dissipative and geometric properties at the different involved scales. Specifically we consider a 'telescopic' system as the one schematized in Fig. 1$_a$ where the dissipation is attained by the friction of the thinner column inside the larger, based on an internal pressure ensuring column friction. The optimization consists in increasing the dissipative energy per unit material volume $\delta$. Despite our mechanism is restricted to one dimensional loading, we argue that the general underlying idea can be extended to more general shapes.

Our optimization scheme takes care of both material strength and columns instability thresholds and it is organized according with the following theoretical scheme. First we optimize the column section geometry and we obtain that the optimal shape is that of an equilateral triangle. Then, we determine the optimal length of superposition and slenderness of the single column. Finally, we extend the optimization to the possibility of selfsimilar decomposition of the column. As we show, the optimality increases with the structure complexity (number of involved length scales) with the theoretical optimal limit attained by an infinite refinement of the system. Interestingly, the resulting fractalization of the transversal section resemble the theoretical fractal Sierpiński triangle construction. Thanks to the simplicity of our scheme all results are analytical with a clear mechanical interpretation. Of course a (technologically and economically) realistic optimal structure is attained when very few refinement scales are considered. We then show that for the proposed scheme already at the third complexity degree the optimal dissipation is almost fully attained (96%). The final optimal dissipation device is then identified based on the possibility of a complete tessellation of the plane by adjacent dissipative columns. As well known, for identical sections this property is possible only for regular polygons with $n = 3, 4, 6$ sides.

It is interesting to remark that the obtained optimal parameters are all of geometrical type and so material independent. Moreover, in the spirit of self-similarity, they are scale-invariant and so may be identically reproduced at the different scales keeping fixed the stress thresholds. As a result the device can have different applications independently from the required dimension. Two other points are important to be mentioned: from one side since our construction is thought to keep the material in the elastic state we may argue that the system can be healed with a simple reversion of the loading sliding; from the other side possible increased properties can be considered, by extending the analysis to hollow sections, but this advantage should be compared to the known technological complications depending on the specific application.

As a final remark we want to point out that we also would like to make some analogies with observed phenomena in natural systems. From one side we refer to the observation of the diffused presence of triangular sections in biological devices, such as triangular shaped indenters as reported *e.g.* in Keten, Xu, and Buehler (2011). From the other side we observe that hierarchical structures are at the base of intriguing optimization of mechanical properties in nature such as in the case of the high dissipative and stiff spider silks (Cranford, Tarakanova, Pugno, & Buehler, 2012). The optimal dissipative properties are attained based on hierarchical structures with contemporary structural optimization at different scales (Wegst & Ashby, 2004). Such complex structures have been recognized as optimal dissipators allowing to attain such macroscopic responses based on materials with standard or poor resistance and stiffness properties (Nosonovsky & Bhushan, 2007). Despite different complex mechanisms may operate or cooperate in these biological examples, we believe that the hierarchical friction mechanism here proposed can shed some light also in the field of optimized dissipative biological materials.

## 2. Optimal dissipative friction column

The proposed dissipation device consists in a planar arrangement of telescopic, friction based, dissipative columns constituted by an hollow cylindrical container and an adhering sliding solid section (see the scheme in Fig. 7). In the following we optimize the geometric properties of the proposed dissipator by searching for the maximum dissipation per assigned volume. As we show this competes with the instability of the compressed sliding column. Observe that, due to the proposed planar tessellation, we may neglect the instability of the hollow containers. We restrict to the hypothesis of *homogeneous, convex, compact sections* of the column, because this leads to the easiest fabric conditions, but extensions of the proposed approach can be easily considered.

### 2.1. Section shape optimization

Consider first the problem of the maximum attainable dissipation for given column weight. The aim is to maximize for the given section area (and column volume) both the force leading to buckling instability and the total friction. To attain the maximum force at fixed area we need to find the shape that corresponds to the smallest value of the slenderness against buckling. First we observe that optimal sections cannot be characterized by different inertia moments for different axes. Indeed, roughly speaking, if $\alpha$ and $\beta$ are the in-plane eigenvectors of the inertia tensor, with $J_\alpha > J_x > J_\beta$ for all directions $x \neq \alpha, \beta$, than one can consider a (fixed area) 'shape variation' (as sketched in Fig. 1b) increasing the minimum inertia moment.

As a result, we may restrict our attention to regular polygon shapes with $n \geq 3$ sides of length $a_n$. Let $r$ be the radius of the circumscribed circle. The polygons can be decomposed in $n$ isosceles triangles as the ones shown in Fig. 1c. The base of the triangle has length

$$a_n = 2r\sin\frac{\pi}{n}, \tag{1}$$

area

$$A_n = r^2 \cos\frac{\pi}{n} \sin\frac{\pi}{n}, \tag{2}$$

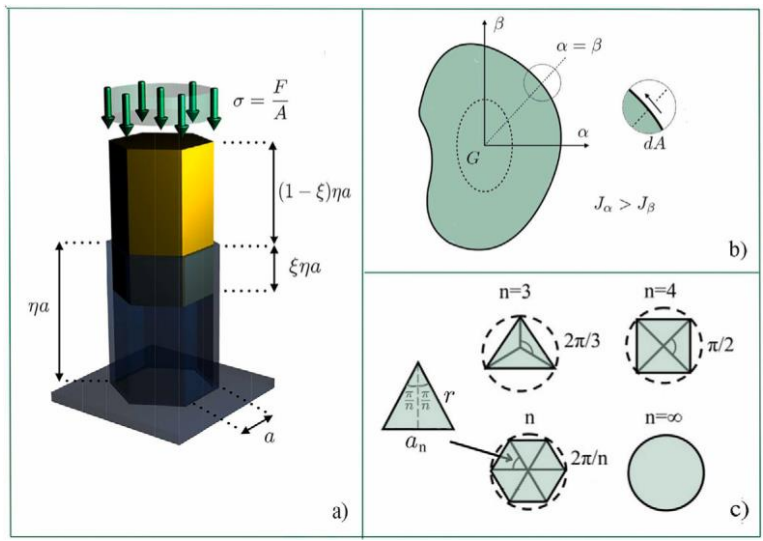

Fig. 1. (a) Scheme of the friction dissipator. (b), (c) shape optimization. In (b) we sketch the reason why regular polygons are optimal with respect to instability, in (c) we show the resulting different possible section shapes.

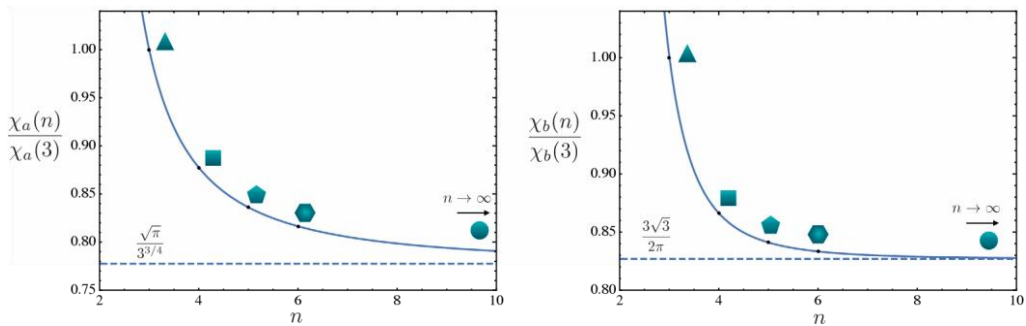

Fig. 2. Representation of the non dimensional optimization parameters as functions of the number of section edges.

and second moment of area with respect to the cylinder axes

$$J_n = \frac{A}{2} r^2 \left( \cos^2\frac{\pi}{n} + \frac{1}{3}\sin^2\frac{\pi}{n} \right). \tag{3}$$

Let then $p = na_n$, $A = nA_n$ and $J = J_{zz} = n/2^n$ be the perimeter, the area, and the in-plane axial inertial moment (that of course for the regular figures is independent from the axis orientation), respectively. We may then introduce the two optimization parameters

$$\chi_a = \chi_a(n) := \sqrt{\frac{p^2}{\pi A}} = 2\sqrt{\frac{n}{\pi}\tan\left(\frac{\pi}{n}\right)} \tag{4}$$

and

$$\chi_b = \chi_b(n) := \frac{J}{A^2} = \frac{J_n}{2nA_n^2} = \frac{1}{4n}\left(\cot\frac{\pi}{n} + \frac{1}{3}\tan\frac{\pi}{n}\right). \tag{5}$$

Observe that to optimize the column we need to increase both parameters. Indeed by increasing $\chi_a$ the lateral surface area at fixed section area grows and thus the friction force for the given sliding friction coefficient increases. On the other hand, by increasing $\chi_b$ we increase the maximum force that can be applied to the column for the assigned area before buckling instability load is attained.

The optimization parameters $\chi_a$ and $\chi_b$ at variable $n$, normalized with respect to the value attained in the case $n = 3$ of the triangular section, are represented in Fig. 2. Observe that the equilateral triangle represents the optimal section shape among the compact convex ones. As a result in the following we restrict our attention to triangular columns. We point out that the obtained geometrical optimization results are material and size-independent. In this respect it may be interesting to remark that the optimality of triangular sections has been observed and described in several compressed biological devices such as the cell-puncture needle of the bacteriophage T4 virus (Keten et al., 2011).

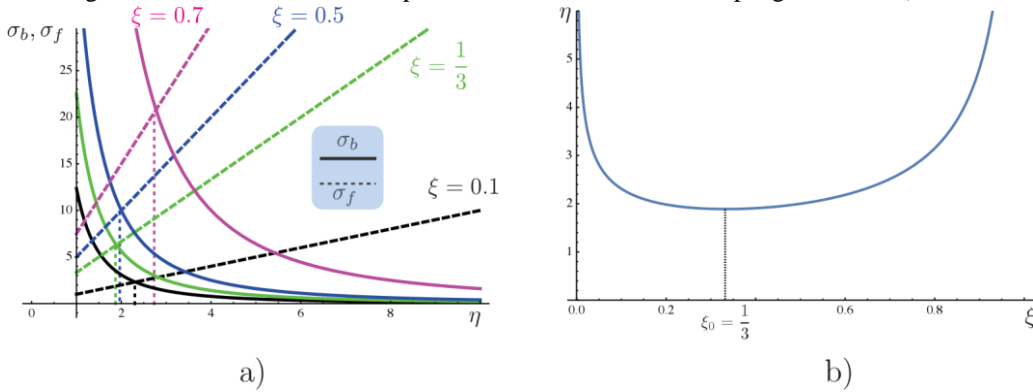

Fig. 3. (a) Dependence of the buckling and friction stress from the slenderness parameter $\eta$ at different value of the insertion fraction $\xi$. (b) representation of the equilibrium condition (10) showing that the lowest value of length and so of the volume is attained at $\xi = \xi_0 = \frac{1}{3}$. Here $c_b = c_f = 10$, but the optimal value $\xi_0 = \frac{1}{3}$ does not depend from these constants.

### 2.2. Height and insertion size optimization

To attain the device volume optimization we require, by controlling the friction stress as described in the following, that the dissipator begins to slide when the stress reaches its limit value corresponding to Euler buckling, i.e. the maximum, attainable load for slender columns. The limit condition of the attainment of the force $F_y = \sigma_y A$ where $\sigma_y$ represents the material limit stress is considered later. It is also important to remark that for simplicity of notation here and in the following we omit the presence of possible safety factors.

Let then $\eta a$ be the height of the column and let $\xi \eta a$ be the insertion length (see Fig. 1). Both the friction force and the Euler buckling force depend on the insertion fraction $\xi$. The total axial force that can be equilibrated by the shear friction stress $\tau$, assumed to be a given constant parameter, is

$$F_f = 3\tau\xi\eta a^2. \tag{6}$$

On the other hand the buckling Euler force $F_b$ for a cantilever loaded at the end – under an assumption of buckling length $L_0 = 2h$, corresponding to a column clamped from one side and free from the other one – and an height $h = (1 - \xi)\eta a$ is

$$F_b = \frac{\pi^2 E}{3} \eta^2 (1-\xi)^2 \frac{a^2}{128} \tag{7}$$

where $E$ is the Young modulus of the material. The corresponding stresses are given by

$$\sigma_f = \hat{\sigma}_f(\xi) = \frac{F_f}{A} = c_f \xi \eta \quad \text{and} \quad \sigma_b = \hat{\sigma}_b(\xi) = \frac{F_b}{A} = c_b \eta^2 (1-\xi)^2, \tag{8}$$

where we introduced the two material parameters

$$c_f = 4\sqrt{3}\tau \quad \text{and} \quad c_b = \frac{\pi^2 E}{96}, \tag{9}$$

depending from the triangular shape through the numerical coefficients. Thus different coefficients would correspond to different section shapes even possible hollow ones. Now to equilibrate the optimal maximum force $F_b$ we need an initial insertion fraction $\xi_0$ such that $F_f = F_b$ that gives

$$\eta = \eta_{eq}(\xi_0) = \sqrt[3]{\frac{c_b}{c_f(1-\xi_0)^2 \xi_0}}. \tag{10}$$

By minimizing $\eta_{eq}$ in (10) with respect to the initial insertion $\xi_0$, we obtain that the minimum value of $\eta$ and thus of volume is attained for

$$\xi_0 = \hat{\xi} = \frac{1}{3}. \tag{11}$$

It is again interesting to point out that the optimal value of 1/3 of inserted height is material and size independent. Moreover it is independent from the section shape because this result does not depend from $c_a$ and $c_b$ (see Fig. 3).

Correspondingly we obtain an optimal slenderness of the beam

$$\hat{\eta} = 3\sqrt[3]{\frac{c_b}{4c_f}} \tag{12}$$

and of the initial equilibrium stress

$$\hat{\sigma} = \hat{\sigma}_f\left(\frac{1}{3}\right) = \left(\frac{1}{3}\right)\sqrt[3]{\frac{c_f^2 c_b}{4}} = \hat{\sigma}_b \leq \sigma_y. \tag{13}$$

Observe that during sliding, by using (8), the 'instantaneous' friction stress is

$$\sigma_f(\xi, \hat{\eta}) = \chi\, c_f\, \xi\, \hat{\eta}, \quad \xi \in [1/3, 1] \tag{14}$$

where we introduced the non dimensional parameter $\chi = \mu_d/\mu_s$, measuring the ratio between the static friction coefficient $\mu_s$ and the dynamic one $\mu_d$. In this respect it is important to observe that

$$\sigma_f(\xi,\hat{\eta}) - \sigma_b(\xi,\hat{\eta}) = \frac{(1-3\xi)^2(3\xi-4)}{9\cdot 2^{2/3}(\xi-1)^2} < 0 \quad \text{for } \xi \neq \frac{1}{3},$$

thus granting that at all values of sliding the system does not buckle.

We may now evaluate the fundamental parameter $\delta$ assigning the maximum attainable *dissipation per unit volume*

$$\delta_v = \int_{1/3}^{1} \sigma_f(\xi)d\xi = \frac{2}{3}\chi\sqrt[3]{2c_f c_b^2} \tag{15}$$

corresponding to a *total dissipation per column*

$$\Delta = \hat{\eta}\frac{3}{4}a^3\, \delta_v = \frac{\sqrt{3}}{2}\chi f_c b^2\, a^3. \tag{16}$$

To develop further the optimization, we may observe that since $c_f$ depends on the shear friction stress $\tau$, we may increase it, varying the slenderness through (13) until the maximum force, corresponding to material failure, is attained

$$\hat{\sigma} = \sigma_y. \tag{17}$$

In other words the system starts to slide in correspondence with the optimal slenderness such the Euler force equals the limit material force $\sigma_y A$. In particular if $\sigma_y$ corresponds to the elastic limit stress this condition grants the possibility of shape recovering after dissipation, an important technological advantage. By (13) and (17) we obtain the corresponding friction stress

$$\tau = \frac{2}{\pi}\sqrt{\frac{2}{E}}\frac{\sigma_y^3}{\gamma} := \frac{\tau}{\sigma_y} = \frac{2}{\pi}\sqrt{\frac{2}{E}}\sigma_y, \tag{18}$$

where we introduced the non dimensional friction parameter $\gamma$. This corresponds to an optimal slenderness of the beam assigned by the height/side ratio

$$\eta = \frac{1}{8}\sqrt{\frac{3}{2}}\pi\sqrt{\frac{E}{\sigma_y}}, \tag{19}$$

and a dissipation per unit volume

$$\delta_v = \frac{4}{3}\chi\sigma_y. \tag{20}$$

To get an analytic measure of the dissipation properties of the attained optimal telescopic device we may consider in the hypothesis of a still device, $\sigma_y = 210$ MPa, $E = 2.1\ 10^5$ MPa, $\mu_s = 0.75$, $\mu_d = 0.54$ so that we obtain the parameters reported in the table in Fig. 4. It is worth noticing that a comparison with a rough measure of the elastic dissipation energy of the material is given by

$$\delta_{mat} \sim \frac{1}{2}\frac{\sigma_y^2}{E} \sim 0.1 \text{J/m}^3 \sim \frac{\delta_v}{2000}.$$

Moreover in the figures we report for comparison the dissipation densities of an industrial dissipator presently adopted in vehicle mechanics.

On the other hand we obtain that for steel

$$\delta_v = 0.96\sigma_y$$

corresponding to the work of a perfectly plastic material with a limit deformation of $\lambda = 1.96$ i.e. with the material reaching a half of its initial length. The variations of the density of dissipation per unit volume $\delta$ and per unit mass $\delta_m = \delta_v/\rho$, where $\rho = 7$ g/cm$^3$ is the steel mass density, are represented for a steel in Fig. 4. Observe also that the corresponding friction force requires a low precompression $\sigma_p = \mu^\tau s = 7.5$ MPa – corresponding to $\sigma_p = \sigma_y/28$ – that can be easily attained during the production process.

It is also interesting to observe that it is possible to consider an active control of the internal pressure with a modulation of the force $F_d$ inducing sliding and dissipation. Then, since during slipping $\xi$ increases from $1/3$ to $1$, the system dissipates without collapsing up to a maximum final force, after all the dissipation, of $F_{max} = 3 F_d$. Finally, it is worth pointing out, that under our assumption that $\hat{\sigma} < \sigma_y$ after dissipation the overall system is still elastic and it can be restored by simply repositioning it into the original configuration $\xi_o = 1/3$.

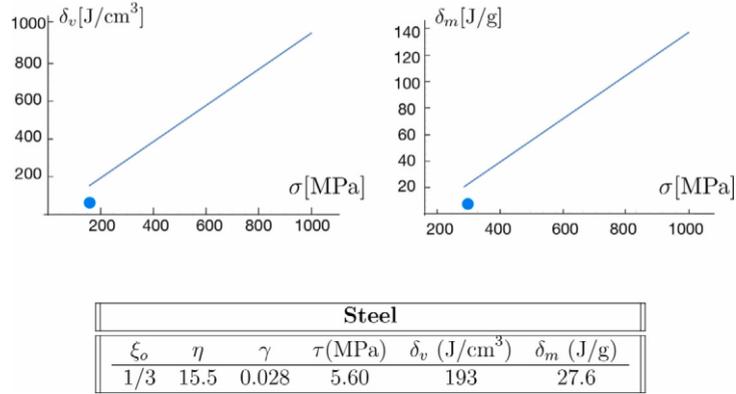

Fig. 4. Optimal parameters for a steel column. With the two dots we report the dissipation densities of an industrial dissipator presently adopted in vehicle mechanics.

## 3. Multiscale optimal dissipative friction columns

In this section, based on the previous observation of size independence of the optimal geometric parameters, to attain an even higher optimization and improve the dissipative behavior of the proposed system, we consider the possibility of a self-similar reproduction of previously described scheme (see Fig. 5). Our geometrical construction consists in dividing the triangular section and the height in three equal parts. The crucial idea at the base of this rescaling is that *the slenderness and the stress corresponding to buckling of the columns are both scale invariant*. This result suggests the possibility of a dissipation mechanism contemporary attained at different length scales. To get simple analytical results we here analyze a simple quasistatic regime, but we point out that a full description of the dynamical behavior of the proposed dissipative device deserves both experimental, numerical, and theoretical deeper investigation that will be the subject of our future work. We remark that the recalled scale-invariance properties have been considered in Puglisi and Truskinovsky (2013) as a possible reason for multiscale self-similar optimization in the evolutionary creation of fibrillar optimal adhesive biological system such as geckoes pads. All the details of the resulting dissipative behavior can be found in Maddalena, Percivale, Puglisi, and Truskinovsky (2009).

The dissipative column of hierarchical levels/complexity $c = 2$ is represented in Fig. 5. Starting by the previously described dissipative column, we assume that the internal slipping pillar is decomposed in a portion ($2/3$ of the total length) of solid section and in a remaining portion ($1/3$ of the total length) shaped to host smaller slipping internal columns with the same slenderness. In this second scale of complexity we have $n = 3^3$ columns with sides of length $a/3$ and height $L/9$. In this way for the smaller columns we have the same values of critical stresses (8) so that the optimal stress undergoing the same compression stress $\hat{\sigma}$ in (13) can be applied. Moreover, based on previous consideration about the independence of the optimal parameters $\xi_o$ in (11) and $\eta$ in (12) we consider the same values. Observe (see Fig. 5) that the system is designed in such a way that three columns out of nine are fixed to the solid section with the other six pillars slipping. It is possible to verify that the average dissipation (per unit volume) is the same

$$\delta_2 = \hat{\eta} a \int_{\frac{1}{3}}^{1} \sigma_d(\xi) \frac{1}{\frac{1}{3}} d\xi = \delta_v. \tag{21}$$

We then deduce that by adding such new system to previous subsystem we have that we may increase the total dissipation by

$$\Delta_2 = \frac{1}{3}\Delta \tag{22}$$

by adding $\frac{2}{9}h$ of beam height. Observe that since we added $1/3$ of the total volume only the total dissipation increases while the dissipation density is constant $\delta_2 = \delta_v$.

It is worth noticing that since $\sigma$ grows with $\xi$ in an identical way at the two considered scales, we may expect in a quasistatic process an identical evolution of the insertion parameter $\xi \in (1/3,1)$, with a cooperative dissipation at both scales. This represents in our opinion an important general property of structural oprimization, leading to an equal distribution of dissipation at different involved scales, that can drive the design of similar hierarchical devivices.

The refinement process can be repeated again by increasing the complexity $c$ and reproducing the reasoning about the scale invariance of the geometrical, stress and dissipation properties (the section of a system with complexity $c = 3$ is represented in Fig. 6). At given complexity $c$ the total initial height attains the value

$$h_{tot}(c) = \sum_{i=1}^{c} \frac{(2)^{i-1}}{9} h \tag{23}$$

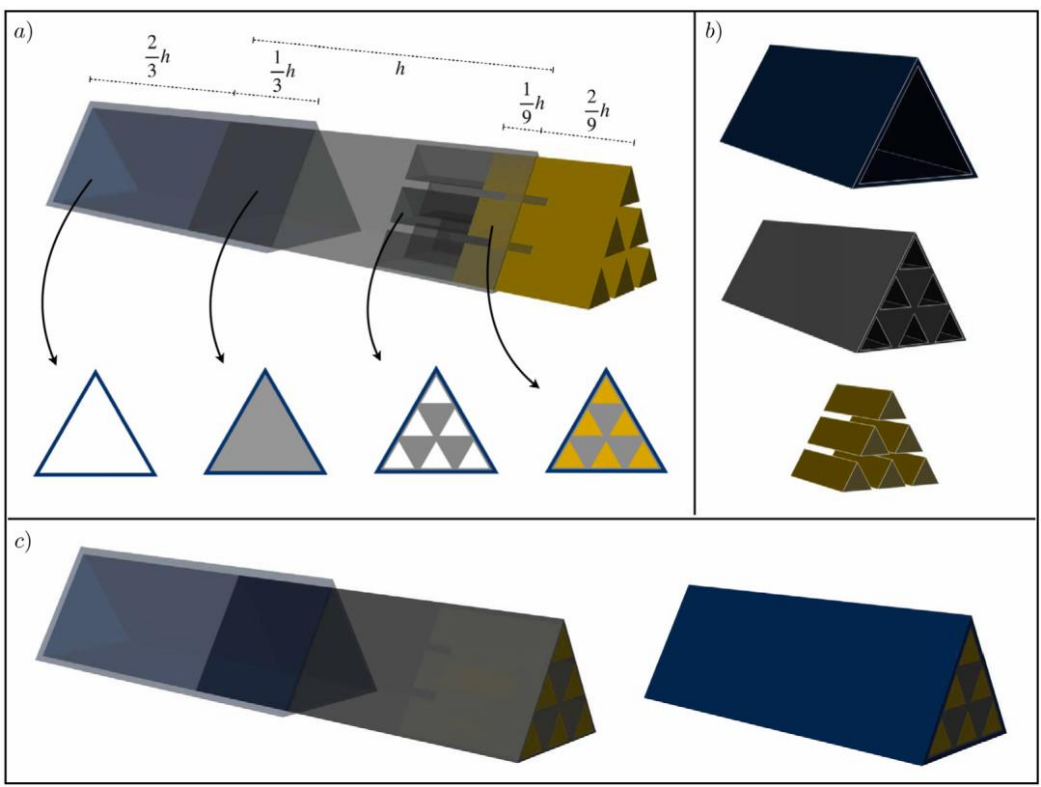

Fig. 5. Scheme of a system with complexity $c=2$. (a) Global view and sections (in the fourth section the three gray triangles are fixed and the yellow ones slip on them), (b) the three composing components, (c) sequential insertion of the two slipping parts.

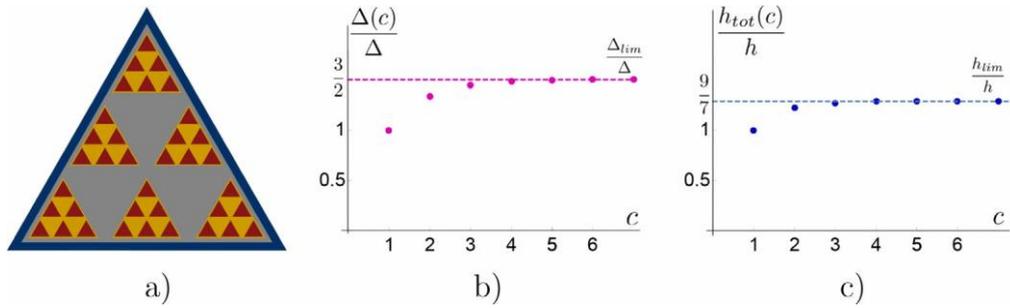

Fig. 6. (a) scheme of the section for a complexity $c=3$; (b), (c) dependence of the total dissipation $\Delta$ and total height $h_{tot}$ on the dissipative device from the complexity $c$.

and the total dissipation is given by

$$\Delta(c) = \sum_{i=1}^{c} \left(\frac{1}{3}\right)^{i-1} \Delta. \tag{24}$$

The section of complexity $c = 3$ is represented in Fig. 6 together with the variation of the total dissipation and height with the complexity. The process can be theoretically increased at any complexity, with a contemporary slide at all the scales and a corresponding increase of total dissipation. The limit length of this theoretical (fractal like) system is

$$\frac{h_{lim}}{h} = \sum_{i=1}^{\infty} \left(\frac{2}{3}\right)^{i-1} = \frac{9}{7}. \tag{25}$$

with a total dissipation

$$\frac{\Delta_{lim}}{\Delta} = \sum_{i=1}^{\infty} \left(\frac{1}{3}\right)^{i-1} = \frac{3}{2}. \tag{26}$$

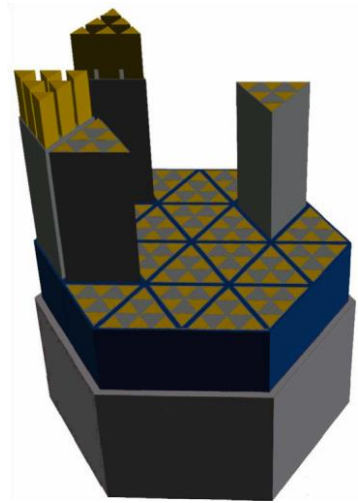

Fig. 7. Planar tessellation with friction based dissipator of complexity three.

Of course this represents an ideal limit, but at a complexity $n = 3$ the optimal complexity is already almost fully attained $\Delta^{(3)}/\Delta = 13/9 \sim 1.44$ corresponding already to the 94% of the limit of infinite refinement. The variations of the dissipation and of the height with the complexity are represented in Fig. 6. Moreover we remark that the self-similarity construction increases the global dissipation $\Delta$ while keeping the dissipation density $\delta$ fixed. It is in any case important to notice that this increase of dissipation is obtained at fixed global section area and so at fixed in plane dissipator size and this can be important in technological application.

The final step of our design of an optimal dissipative device is represented in Fig. 7. It is worth noticing that as well known the plane can be tessellated by identical figures only in the special cases of equilateral triangles, hexagons or squares. As a result our choice of regular triangles represents an optimal shape also in this perspective. Indeed we can maximize for the given area of the dissipation device the area covered by dissipative columns. Importantly, this is also important in the perspective that, as we anticipated, the choice of a planar tessellation avoids buckling instabilization of the containers at the highest dimension scale.

## 4. Discussion

We propose a new telescopic hierarchical dissipation device optimized in both the shape and complexity by considering the possibility of self-similarity and contemporary dissipation at different length scales. This dissipative device starts dissipating at a given design force and with the possibility of sustaining without collapsing a final force, after dissipation, that is three times the initial force. We began by optimizing the dissipation properties for the single column with a resulting behavior summarized in Fig. 4. By choosing the dimension and the number of dissipators the force $F_{el}$ that switches on the dissipative process can be chosen. After dissipation begins, the system is able to equilibrate a force $F_{el} < F < 3F_{el}$. It is very interesting to notice that the comparison of the dissipation density $\delta_v$ with the material dissipation density $\delta_{el} \sim \frac{1}{2} \frac{\sigma_E^y}{E}$ shows the great advantage of the proposed friction based dissipator: $\delta_v \sim 2000 \delta_{el}$. This dissipation is not attained by plastic dissipation, but by the sliding friction work of concentric columns. As a result, we can argue that after dissipation the device can be restored to the initial configuration repositioning previously slided pillars in the original configuration based on appropriate lubrication systems. The behavior has then been improved, by considering a hierarchical reproduction of the optimized behavior of the single column dissipator. As we show, the device can have increasing degree of complexity with the interesting aspect that the system contemporary dissipates at all the involved scales, thus optimizing the dissipation process based on the scale invariance of the Euler instability stress. The total dissipation can be increased of 50% increasing the device height of only 2/7. While this value is only theoretical, because it refers to the limit fractal system, the maximum dissipation value is almost attained already at a complexity $c = 3$ (94% of the theoretical limit). Interestingly the selfsimilar system has a section resembling the theoretical fractal Sierpiński triangle construction. Loosely speaking, the described properties have some analogy with biological materials that typically owe their special material properties to their low scale geometry and to their hierarchical character and not to the properties of the composing materials (Huang et al., 2019). Indeed, as we show, the dissipation per unit volume is much higher than the dissipation that one could obtain by plasticizing the material.

Several interesting augmentations of the model can be considered. From one side there is the possibility of actively controlling the friction based on the internal pressure, thus obtaining a controllable dissipation system that can be activated at variable force. Moreover, it is possible to consider variable friction at different scales, with corresponding variation of the length accordingly to previous treatment, allowing differential slipping at different forces. Thus for small forces only the smaller columns could be activated, whereas for larger forces the number of involved scales could grow. A corresponding optimization for dynamical dissipation and large spectra attenuation could be considered (Miniaci et al., 2018). All these properties suggest that the new proposed concept for dissipation can be of interest in different directions of applications.

Declaration of competing interest

The authors declare that they have no known competing financial interests or personal relationships that could have appeared to influence the work reported in this paper.

Acknowledgments

GP has been supported by the Italian Ministry of Education, University and Research (MIUR) PRIN project ''Mathematics of active materials: From mechanobiology to smart devices'' (2017KL4EF3) and by GNFM (INdAM). NMP is supported by the European Commission under the FET Open ''Boheme'' grant no. 863179, as well as by the MIUR, Italy under the PRIN-20177TTP3S.